\documentclass[article]{emulateapj}
\slugcomment{Draft}

\shorttitle{Solar Composition from Genesis Data}
\shortauthors{Laming et al.}

\begin{document}

\title{Determining the Elemental and Isotopic Composition of the preSolar Nebula
from Genesis Data Analysis: The Case of Oxygen}


\author{J. Martin Laming\altaffilmark1, V. S. Heber\altaffilmark2, D. S. Burnett\altaffilmark3, Y. Guan\altaffilmark3,
R. Hervig\altaffilmark4, G. R. Huss\altaffilmark5, A. J. G.
Jurewicz\altaffilmark4, E. C. Koeman-Shields\altaffilmark5, K. D.
McKeegan\altaffilmark2, L. Nittler\altaffilmark6, D. B.
Reisenfeld\altaffilmark7, K. D. Rieck\altaffilmark8, J. Wang\altaffilmark6,
R. C. Wiens\altaffilmark8, \& D. S. Woolum\altaffilmark9}


\altaffiltext{1}{Space Science Division, Naval Research Laboratory, Code
7684, Washington DC 20375 \email{laming@nrl.navy.mil}} \altaffiltext{2}{Dept.
of Earth, Planetary \& Space Sciences UCLA, Los Angeles CA 90095}
\altaffiltext{3}{Div. of Geological \& Planetary Sciences, Caltech, Pasadena
CA 91125} \altaffiltext{4}{School of Earth \& Space Exploration, Arizona
State University, Tempe, AZ 85287} \altaffiltext{5}{Hawaii Institute of
Geophysics \& Planetology, University of Hawaii at Manoa, 1680 East-West
Road, Honolulu, HI 96822} \altaffiltext{6}{Department of Terrestrial
Magnetism, Carnegie Institute of Washington, Washington DC 20015}
\altaffiltext{7}{Department of Physics, University of Montana, Missoula, MT
59812} \altaffiltext{8}{Space and Remote Sensing (ISR-2), LANL, Los Alamos NM
877545}\altaffiltext{9}{Department of Physics, CSUF, Fullerton CA 92831}

\begin{abstract}
We compare element and isotopic fractionations measured in solar wind
samples collected by NASA's Genesis mission with those predicted from
models incorporating both the ponderomotive force in the chromosphere and
conservation of the first adiabatic invariant in the low corona. Generally
good agreement is found, suggesting that these factors are consistent with
the process of solar wind fractionation. Based on bulk wind measurements,
we also consider in more detail the isotopic and elemental abundances of O.
We find mild support for an O abundance in the range 8.75 - 8.83, with a
value as low as 8.69 disfavored. A stronger conclusion must await solar
wind regime specific measurements from the Genesis samples.

\end{abstract}

\section{Introduction}
Solar system bodies formed from the pre-solar nebula, but at different
places, at different times and through different processes. Variations in
their elemental and isotopic compositions observed today give clues to the
mechanisms of formation of these different bodies. A major problem has been
our lack of knowledge of the original composition of the solar nebula.
Although the Sun represents 99.86\% of the known mass of the solar system,
its elemental composition revealed by remotely sensed spectroscopy of its
photosphere is not determined with sufficient precision to meet planetary
science needs, and its isotopic composition hardly known at all.

NASA$^{\prime}$s Genesis mission \citep{burnett13,burnett17} was designed to
solve these problems by collecting samples of solar wind which were then
returned to Earth for analysis in laboratory mass spectrometers at far higher
precision and better calibration than can be achieved in flight. Genesis
orbited the L1 Lagrange Point between 2001 December 3 and 2004 April 1
collecting solar wind ions in various different collector materials. Despite
the setback caused by the crash of the Sample Return Capsule upon return to
Earth, high accuracy element abundance results now exist for bulk solar wind
samples for over a dozen elements. Additionally, isotopic abundances have
been measured in the bulk solar wind for N, O, He, Ne, Ar, Kr, and Xe, and
isotopic fractionation between fast and slow solar wind regimes has been
measured for a subset of these elements (He, Ne, Ar).

This suite of data represents an opportunity to compare precise and accurate
solar wind composition with that of the underlying solar composition.
Elemental fractionation between the solar photosphere and corona and wind has
been known since 1963 \citep{pottasch63}. Elements with first ionization
potential (FIP) below about 10 eV (e.g. Mg, Si, Fe; those that are
predominantly ionized in the solar chromosphere) are seen to be enhanced in
abundance in the corona by a factor of about 3-4 relative to the so-called
high FIP elements (e.g. H, O, Ar) which are mainly neutral below the corona.
Similar fractionation is seen in the solar wind, although it varies with
solar wind regime; the fast wind being less fractionated in this manner than
the slow speed wind \citep[e.g.][]{bochsler07,pilleri15}

This FIP fractionation is now understood as being due to the action of the
ponderomotive force \citep{laming04,laming09,laming12,laming15,laming16}.
This arises as magnetohydrodynamic (MHD) waves propagate through, or reflect
from the solar chromosphere. If, as recent observations suggest
\citep[e.g.][]{depontieu07}, these waves carry significant energy and
momentum in the solar atmosphere, then any change in their direction of
propagation due to density gradients in the Sun must result in a net force on
the plasma. Since the waves of interest here are fundamentally oscillations
of the magnetic field (Alfv\'en and fast mode waves, collectively known as
``Alfv\'enic'' when close to parallel propagation), they only interact with
the ionized fraction of the plasma. Hence the ponderomotive force separates
ions from neutrals.

\begin{figure*}[t]
\centerline{\includegraphics[scale=0.45]{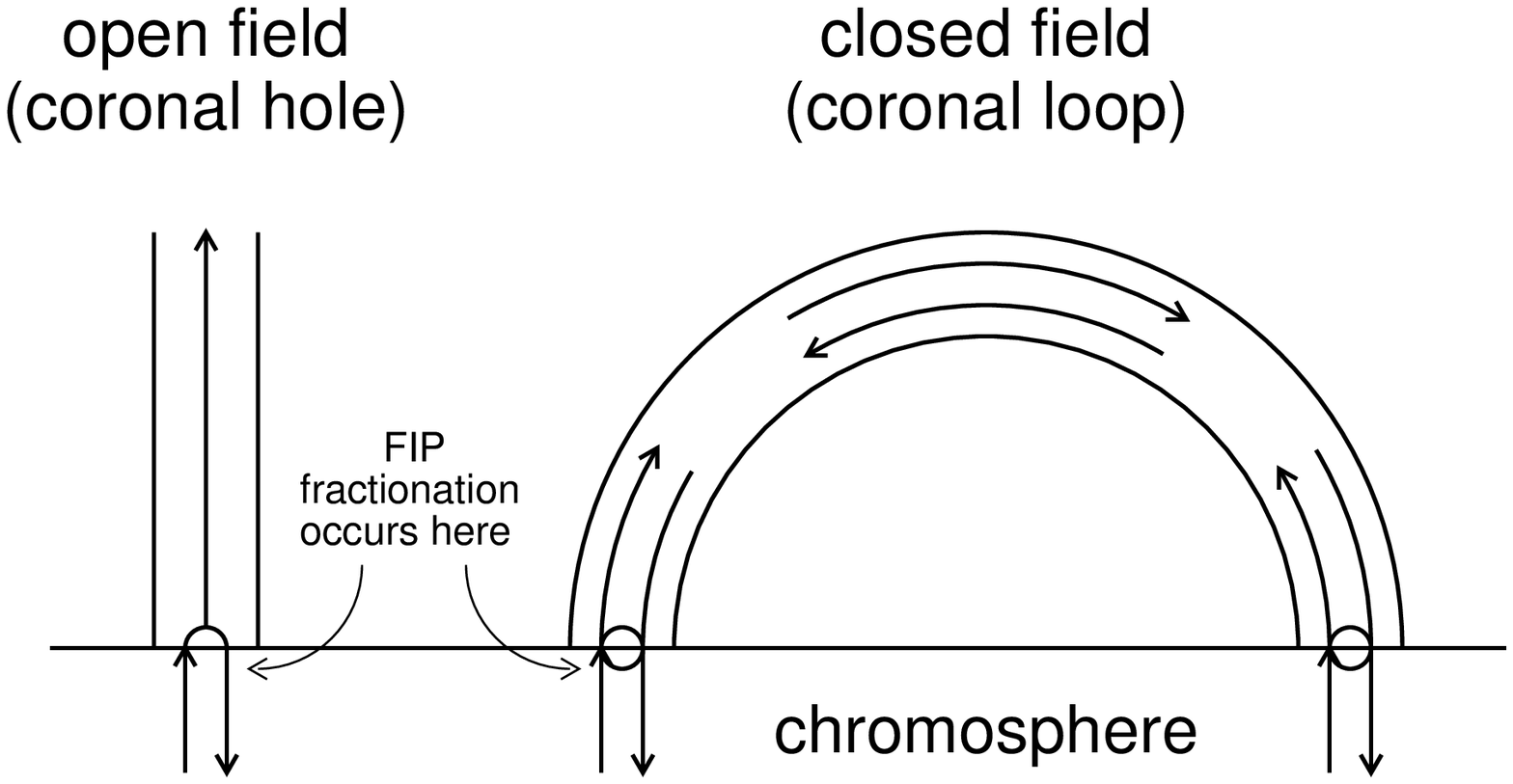}\includegraphics[scale=0.45]{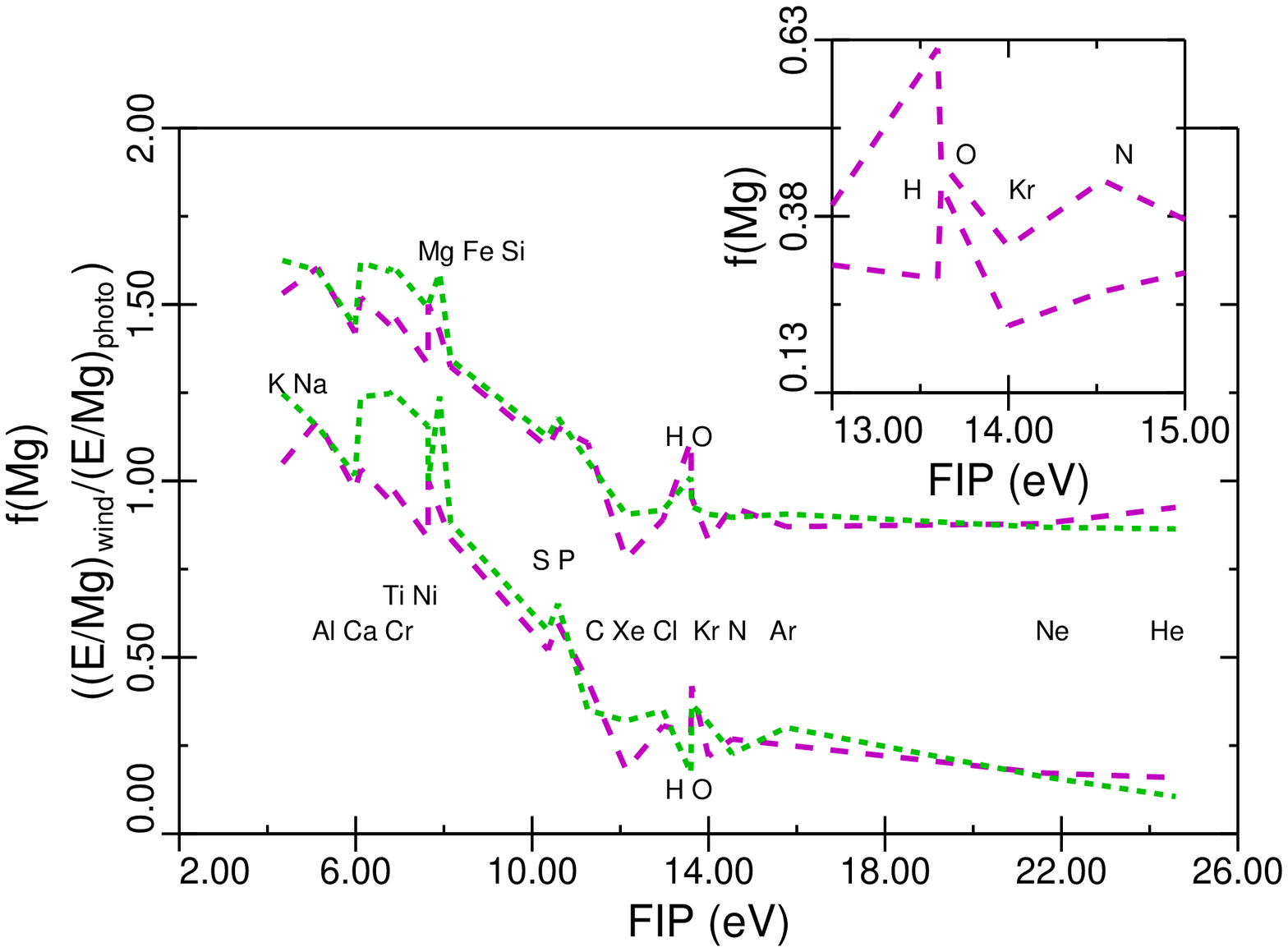}}
\caption{\label{fig1} Left: Schematic showing FIP fractionation in open and closed field
regions. In both cases, waves impinge on footpoints from below, but the closed field
can also have wave generation within the corona. Right: Model element fractionation in
open (top, shifted up by 0.5 for clarity) and closed field (bottom). The result shown here is for
model 1 (see Table 1). Green short dash lines show
ponderomotive acceleration only; purple long dash lines show combined effect of ponderomotive
acceleration and adiabatic invariant conservation. The inset shows the region around H, O, Kr, and N
in more detail, for fast wind (top) and slow wind (bottom), for the combined model only.
Note that the fractionation ratio O/H $>1$ in the slow wind, but is $<1$ in the fast wind.}
\end{figure*}

The FIP fractionation, including the depletion of He, is most faithfully
reproduced in a model of a closed coronal loop where the Alfv\'en waves are
resonant \citep{laming12,laming16,rakowski12}, so that the coronal loop acts
as a resonant cavity, where the Alfv\'en wave travel time from one footpoint
to the other is an integral number of wave half periods. Although it is
possible for waves ultimately deriving from convection within the solar
envelope to enter coronal loops at footpoints and propagate into the corona,
typically the periods of these waves (three or five minutes) are too long for
resonance. Resonant waves are most plausibly excited within the coronal loop
itself, most likely as a byproduct of the mechanism(s) that heat the corona
\citep{dahlburg16}. In open field regions, such a resonance does not exist,
and only waves propagating up from footpoints are possible. In such a
scenario, the difference in fractionation between fast wind which originates
in open magnetic field structures on the Sun, and slow wind which originates
in closed coronal loops which are subsequently opened up by interchange
reconnection \citep[e.g.][]{lynch14}, arises naturally due to the extra
resonant waves. Figure 1 (left panel) gives a schematic illustration of the
open and closed field models, and the right panel illustrates the different
fractionation patterns (see below for fuller discussion).

\section{Model Calculations}
The fractionation is calculated in each case by solving Alfv\'en wave
transport equations in a model coronal structure. In the open field region a
spectrum of Alfv\'en waves is chosen to match those given in
\citet{cranmer05} and \citet{cranmer07} high up in the corona, and integrated
back to the chromosphere. In the closed loop, we take a single Alfv\'en wave
corresponding to the fundamental of a 75,000 km long loop having a 10 G
coronal magnetic field, combined with two additional photospheric waves with
periods of three and five minutes \citep[e.g.][]{heggland11}. All waves are
taken to be shear (planar) Alfv\'en waves \citep{laming16}. The instantaneous
ponderomotive acceleration, $a$, is given by
\begin{equation}
a={c^2\over 2}{\partial\over\partial z}\left(\delta E^2\over B^2\right)
\end{equation}
where $\delta E$ is the wave electric field, $B$ the ambient magnetic field,
$c$ the speed of light, and $z$ is a coordinate along the magnetic field. The
element fractionation, $f_p$, is calculated from ratios of densities $\rho
_k$ for element $k$ at upper and lower boundaries of the fractionation region
$z_u$ and $z_l$ respectively, as given by the equation \citep{laming16}
\begin{eqnarray}
\nonumber f_p&=&{\rho _k\left(z_u\right)\over\rho _k\left(z_l\right)}\\ &=&\exp\left\{
\int _{z_l}^{z_u}{2\xi _ka\nu _{kn}/\left[\xi _k\nu
_{kn} +\left(1-\xi _k\right)\nu _{ki}\right]\over 2k_{\rm B}T/m_k+v_{||,osc}^2+2u_k^2}dz\right\},
\end{eqnarray}
where $\xi _k$ is the element ionization fraction, $\nu _{ki}$ and $\nu
_{kn}$ are collision frequencies of ions and neutrals with the background gas
(mainly hydrogen and protons), $k_{\rm B}T/m_k \left( =v_z^2\right)$
represents the square of the element thermal velocity along the
$z$-direction, $u_k$ is the upward flow speed and $v_{||,osc}$ a longitudinal
oscillatory speed, corresponding to upward and downward propagating sound
waves. Because $\nu _{ki}>>\nu _{kn}$ in the fractionation region at the top
of the chromosphere, small departures of $\xi _k$ from unity can result in
large decreases in the fractionation.

\begin{figure*}[t]
\centerline{\includegraphics[scale=1.0]{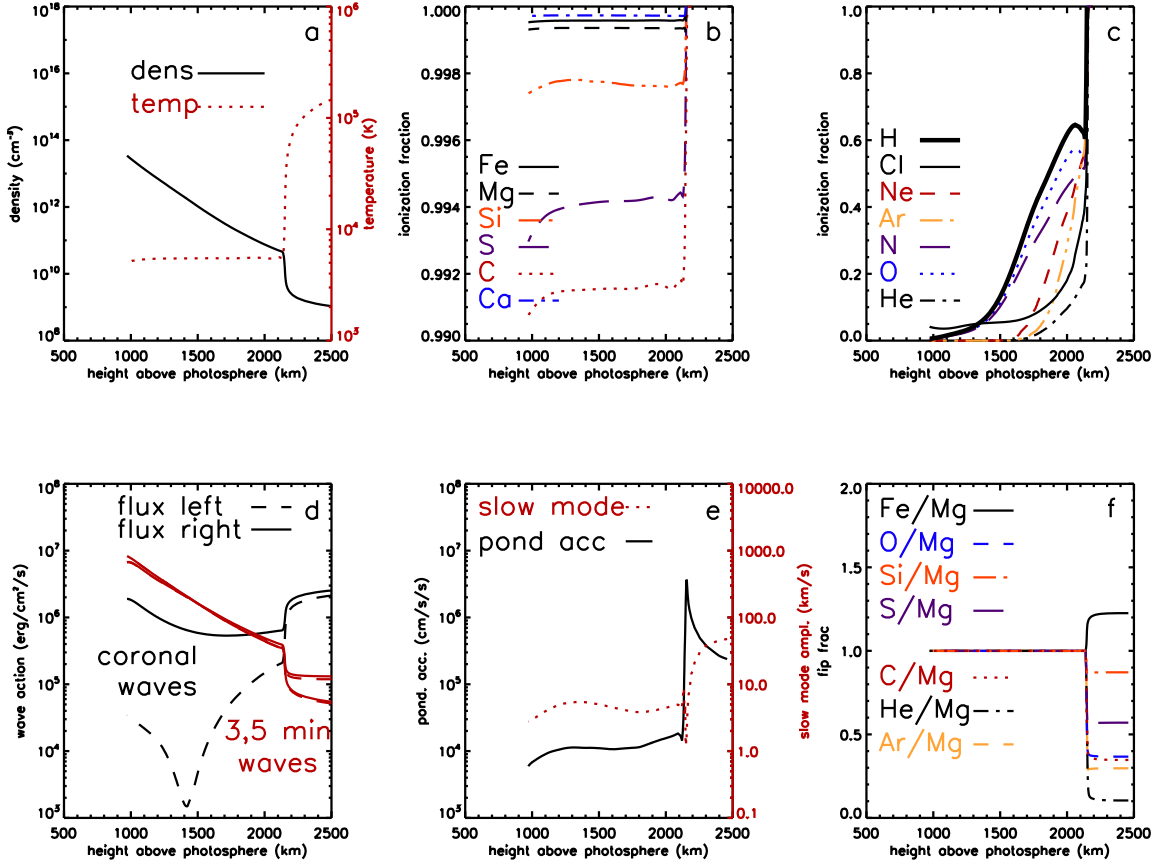}}
\caption{\label{fig2}The chromospheric model. (a) shows the
density and temperature structure of the chromosphere. (b) shows chromospheric
ionization fractions for low FIP elements and (c) for high FIP elements.
(d) shows the wave energy fluxes in each direction for the three waves
in the closed loop model. (e) shows the ponderomotive acceleration (solid line)
and the
amplitude of slow mode waves induced by the Alfv\'en wave driver. (f) shows the
fractionations resulting for selected elements relative to Mg. Gas pressure and magnetic field
pressure are equal at about 1000 km, magnetic field pressure dominating at higher altitudes.}
\end{figure*}

Isotopic fractionation between fast and slow solar wind has also been
observed in the Genesis data. Specifically, lighter isotopes are more
abundant relative to heavy ones of the same element in the slow wind compared
to the fast \citep{heber12}. This is the opposite of what would be expected
from equation 2, where with increased ponderomotive acceleration, $a$, a
heavier isotope would have a smaller thermal speed and hence a higher value
of $f_p$. An extra mass dependent fractionation (MDF) mechanism must be
present. Inefficient Coulomb drag (ICD) has frequently been discussed,
especially in connection with the depletion of He in the solar wind
\citep{bodmer98,bochsler07b}. This depletion is now part of the FIP
fractionation. During the Genesis data collection period, there is little
other evidence for ICD in data collected by Genesis, Wind \citep{kasper12},
or the Advanced Composition Explorer \citep[ACE;][]{pilleri15} in element
abundances (the solar minimum of 2007-8 might be a different matter). ICD
should be strongest in the fast wind emanating along open field lines in
coronal holes, with slow wind originating in closed loops more fully mixed by
waves and turbulence; the opposite of what we see.

We argue therefore that the MDF of isotopes is most likely due to the
conservation of the first adiabatic invariant, in conditions where the ion
gyrofrequency $\Omega =eB/m_kc >> 1/\tau _{coll}
>> v_{ex}/R$. The first inequality means that an ion executes many
gyro-orbits around the magnetic field line in the time between Coulomb
collisions with other ions, $\tau _{coll}$, and thus the magnetic flux
enclosed by its orbit is conserved. Hence $Br_g^2\propto
\left(v_x^2+v_y^2\right)/B=v_{\perp}^2/B$ is constant ($r_g$ is the particle
gyroradius), giving rise to an acceleration
\begin{equation}
{dv_z\over dt}=-{1\over 2}{dB\over dz}{v_{\perp}^2\over B}
\end{equation}
in conditions where $v^2=v_z^2+v_{\perp}^2$ is constant. The second
inequality expresses the condition that the plasma remain otherwise {\em
collisional}, in that Coulomb collision frequencies are much greater than the
expansion rate (wind speed, $v_{ex}$, divided by radius, $R$) of the solar
wind, and local abundance enhancements in the corona can be sustained by
increased diffusion up from the solar photosphere. This is necessarily a
loose concept, and so our approach is to calculate the FIP fractionation for
open and closed field according to the models outlined above, and then the
add in mass dependent fractionation (which arises because the thermal speeds
$v_{\perp}^2$ and $2k_{\rm B}T/m_k$ are proportional to $1/m_k$, while
$v_{||,osc}^2$ and $u_k^2$ representing fluid motions are not, and are
usually much larger)
\begin{equation}
f_a=\exp\left\{ -\int {{dB/dz}\left(v_{\perp}^2/B\right)\over
2k_{\rm B}T/m_k+v_{||,osc}^2+2u_k^2}dz\right\},
\end{equation}
to match the isotopic differences between high speed and low speed solar
wind. The region of integration in equation 4 is in the corona, out to a
heliocentric distance of $1.5 - 2 R_{\Sun}$, where the corona is sufficiently
collisionless to allow solar wind acceleration to commence \citep{cranmer99,
miralles01}. Figure 1 (right panel) shows the resulting fractionations
relative to Mg for open (top curves, shifted upwards by 0.5 for clarity) and
closed field (bottom). The green lines indicate the effect of the
ponderomotive acceleration only, the purple curves show the combined effect
of ponderomotive acceleration and the adiabatic invariant conservation.
Elements lighter (heavier) than Mg are enhanced (depleted) in abundance by
the adiabatic invariant.

Figure 2 illustrates some important features of the FIP fractionation in
closed loops, based on the chromospheric model of \citet{avrett08}. Top left
(a) shows the density and temperature structure of the chromosphere. Top
middle (b) shows chromospheric ionization fractions for low FIP elements, and
top right (c) for high FIP elements. Bottom left (d) shows the wave energy
fluxes in each direction for the three wave frequencies considered, the wave
resonant with the coronal loop, and three and five minute waves propagating
up from the photosphere. Bottom middle (e) shows the ponderomotive
acceleration (solid line) and the amplitude of slow mode waves induced by the
Alfv\'en wave driver. Bottom right (f) shows the fractionations resulting for
selected elements relative to Mg. The ponderomotive acceleration has a strong
``spike'' at an altitude of 2150 km, where the chromospheric density gradient
is steep (see top left), resulting in strong fractionation at this height.

\begin{table*}
\begin{center}

\caption{Isotopic Fractionations}

\begin{tabular*}{\textwidth}{|c @{\extracolsep{\fill}} |ccc|}
\hline
Ratio & Model 1 (low MDF)& Model 2 (high MDF)& Observations\\
\hline  
$^3$He/$^4$He& -4.6\%& -5.3\%& $6.31\pm 0.21$\% $^1$   \\
$^{20}$Ne/$^{22}$Ne& 0.46\% amu$^{-1}$& 0.41\% amu$^{-1}$& $0.42\pm 0.05$\% amu$^{-1}$ $^1$\\
$^{36}$Ar/$^{38}$Ar& 0.25\% amu$^{-1}$& 0.20\% amu$^{-1}$& $0.26\pm 0.05$\% amu$^{-1}$ $^1$\\
\hline
$f_{FIP, slow}$ & 2.69 & 2.73& 2.65$^2$\\
$f_{FIP, fast}$ & 1.91 & 1.99& 2.03$^2$\\
\hline
$B_{freeze, slow}/B_{\Sun}$ & 0.135$^3$ & 0.105$^3$& 0.094$^4$\\
$B_{freeze, fast}/B_{\Sun}$ & 0.368$^3$ & 0.235$^3$& 0.173$^4$\\

\hline
\end{tabular*}

\end{center}
{\tablecomments{$^1$ data from Heber et al. (2012a), slow wind relative to
fast wind; $^2$ Pilleri et al. (2015). $B$ field expansions $^3$ are adjusted
to give the best fit to the Ne and Ar isotpic ratios, and $^4$ estimated from
Wang \& Sheeley (1990).} \label{tab1}}
\end{table*}

\section{Results \& Discussion}
We compare the measured fractionations from Genesis samples with models
designed to match the solar wind conditions during the Genesis period, and
seek a ``best fit''. In this paper, as a short cut, we construct individual
fast and slow wind models (including the adiabatic invariant), given above in
Figure 1 (right panel). These have been tuned to match the observed FIP
fractionations given by \citet{pilleri15}, defined as the sum of the FIP
fractionations for Fe, Mg, and Si divided by the sum of those for C, O, and
Ne. We assume a time fraction 0.35 during this time period due to fast wind,
and 0.65 from slow wind and coronal mass ejections (CMEs), assumed to be
similarly fractionated \citep{pilleri15}. This then matches the ratio of Mg
fluences in fast and slow wind/CMEs,
$\left(0.35f_{FIPfast}\right)/\left(0.65f_{FIPslow}\right)$ given by
\citet{heber14}.

Further details of these models are given in Table 1. The assumed diminution
of magnetic field, which controls the adiabatic invariant acceleration, is
compared to that estimated from \citet{wang90}. These authors give values for
$B_s\left(R_s\right)$ at $R_s=2.5R_{\Sun}$ relative to its value on the solar
surface. We estimate the magnetic field decrease at $1.5 - 2.0 R_{\sun}$
where the solar wind decouples collisonally from the sun to be approximately
$\sqrt{B_S\left(R_S=2.5 R_{\sun}\right)/B_{\sun}}$ and compare this with our
assumed model values in Table 1. We assume representative speeds of 450 and
600 km s$^{-1}$ for slow and fast wind respectively \citep{pilleri15}. We
emphasize that this magnetic field decrease represents the least constrained
free parameter in the model, and is chosen to match existing solar
observations, and in combination with the FIP fractionation reproduce data
from both ACE and Genesis simultaneously.

We give two models with differing amounts of mass dependent fractionation
(MDF) corresponding to different magnetic field expansions,
$B_{freeze}/B_{\sun}$, yielding different isotopic fractionations. Both
models have been specified to reproduce the observed fractionation between
fast and slow wind in $^{20}$Ne/$^{22}$Ne and $^{36}$Ar/$^{38}$Ar, as given
in \citet{heber12}. The ratio $^3$He/$^4$He shows similar behaviour that is
not accounted for, due to other by now well known processes involving the
resonant absorption of ion-cyclotron waves that arises for $^3He$ alone
because of its unusual charge to mass ratio of 2/3. These enhance the $^3$He
abundance \citep[e.g.][]{bucik14} and are currently not included in our
model. Table 1 shows that the adjusted model slow-fast wind difference in
both Ne and Ar isotopic compositions match well with Genesis data.

In Table 2 we compare isotopic fractionations derived by application of
models 1 and 2 to the Genesis results with previous inferences in the
literature. The modeled fractionations of $^{14}$N/$^{15}$N,
$^{16}$O/$^{18}$O and $^{25}$Mg/$^{26}$Mg are given for the combined fast and
slow, i.e. bulk solar wind observed by Genesis, and compared with
observations where they exist. Agreement is quite good, with the Sun
isotopically lighter than other solar system bodies \citep[c.f.][]{ayres13}.
By combining our N model fractionations with the Genesis solar wind
$^{14}$N/$^{15}$N of Marty et al. (2011) we calculate photospheric
$^{14}$N/$^{15}$N ratios (Table 2) which can be compared with values for
Jupiter and Saturn, often presumed to have formed from the the same pre-solar
nebula material accreting the same N$_2$ as the Sun.

Figure 3 shows the predicted elemental fractionation for bulk (i.e. time
integrated) solar wind collected by Genesis. The left and right panels give
results for models 1 and 2 as given in Table 1, which have lesser and greater
degrees of mass dependent fractionation by conservation of the first
adiabatic invariant respectively. The two models give very similar FIP plots.
The symbols in Figure 3 (same in both panels) give the measured Genesis
fractionations relative to the photospheric abundances of \citet{asplund09},
\citet{scott15a,scott15b} and \citet{grevesse15}. The Genesis results are K,
Na, \citet{rieck16}; Ca, Al, Cr, \citet{heber14}; Fe, Mg, \citet{jurewicz11};
C, N, O, \citet{heber13}; Kr, Xe, \citet{meshik14}; and H,
\citet{koeman-shields16}.

The overall agreement between theory and data on Figure 3 is quite good.
Inclusion of the adiabatic invariant is a non-negotiable part of the model;
it is required to provide the good matches in isotopic ratios shown in Table
1.  Exclusion of the adiabatic invariant makes little difference for high FIP
elements in Figure 3. The results excluding the adiabatic invariant better
match the magnitude of the observed f(Mg) in Figure 3 for the low FIP range
between Na and Mg; however the low FIP trend of the Genesis data is better
matched by including the adiabatic invariant but the predicted magnitudes low
FIP F(Mg) between Na and Mg are slightly too low relative to the data.  Both
models predict a small Fe/Mg fractionation that is not present in the data.

\begin{figure*}[t]
\centerline{\includegraphics[scale=0.45]{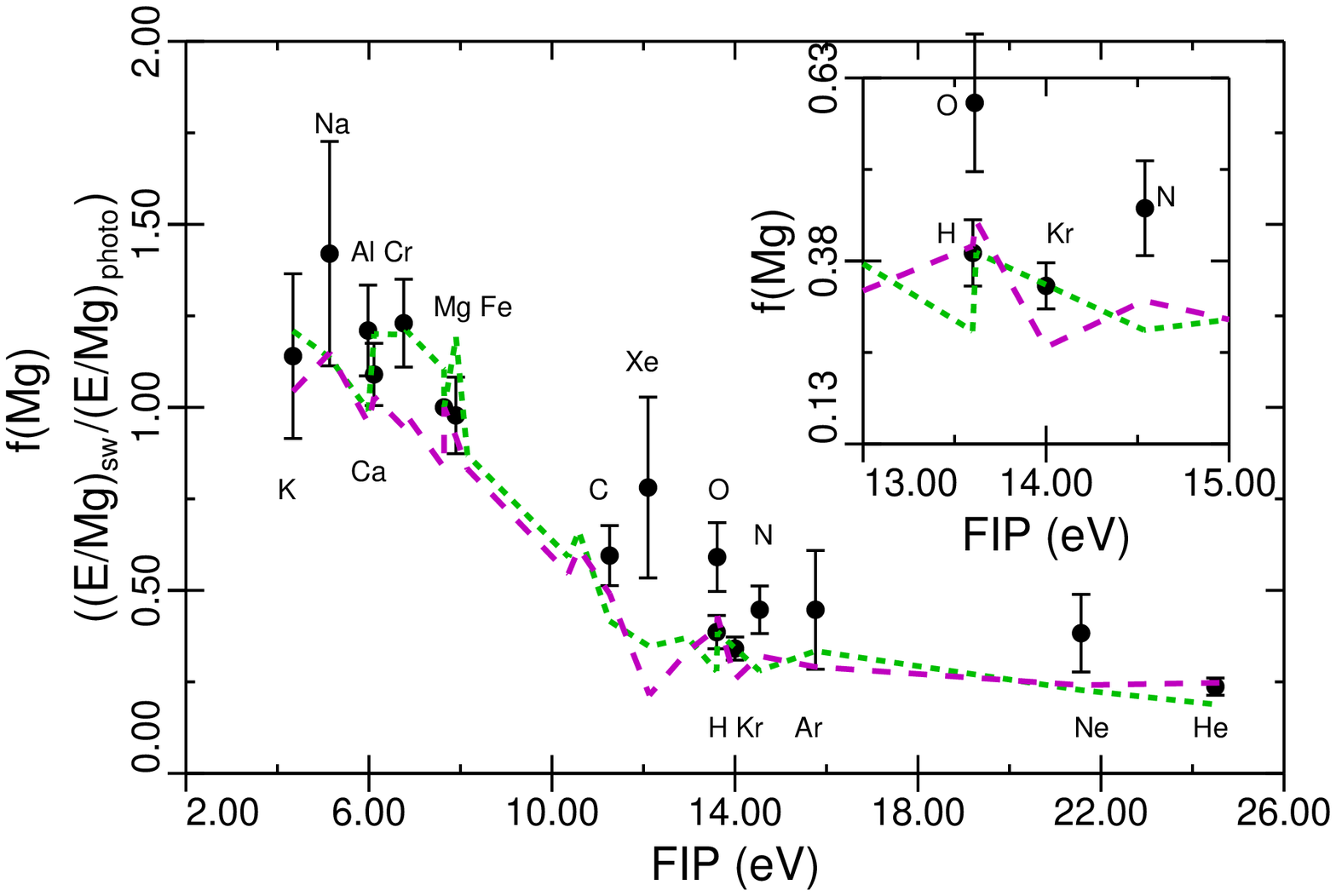}\includegraphics[scale=0.45]{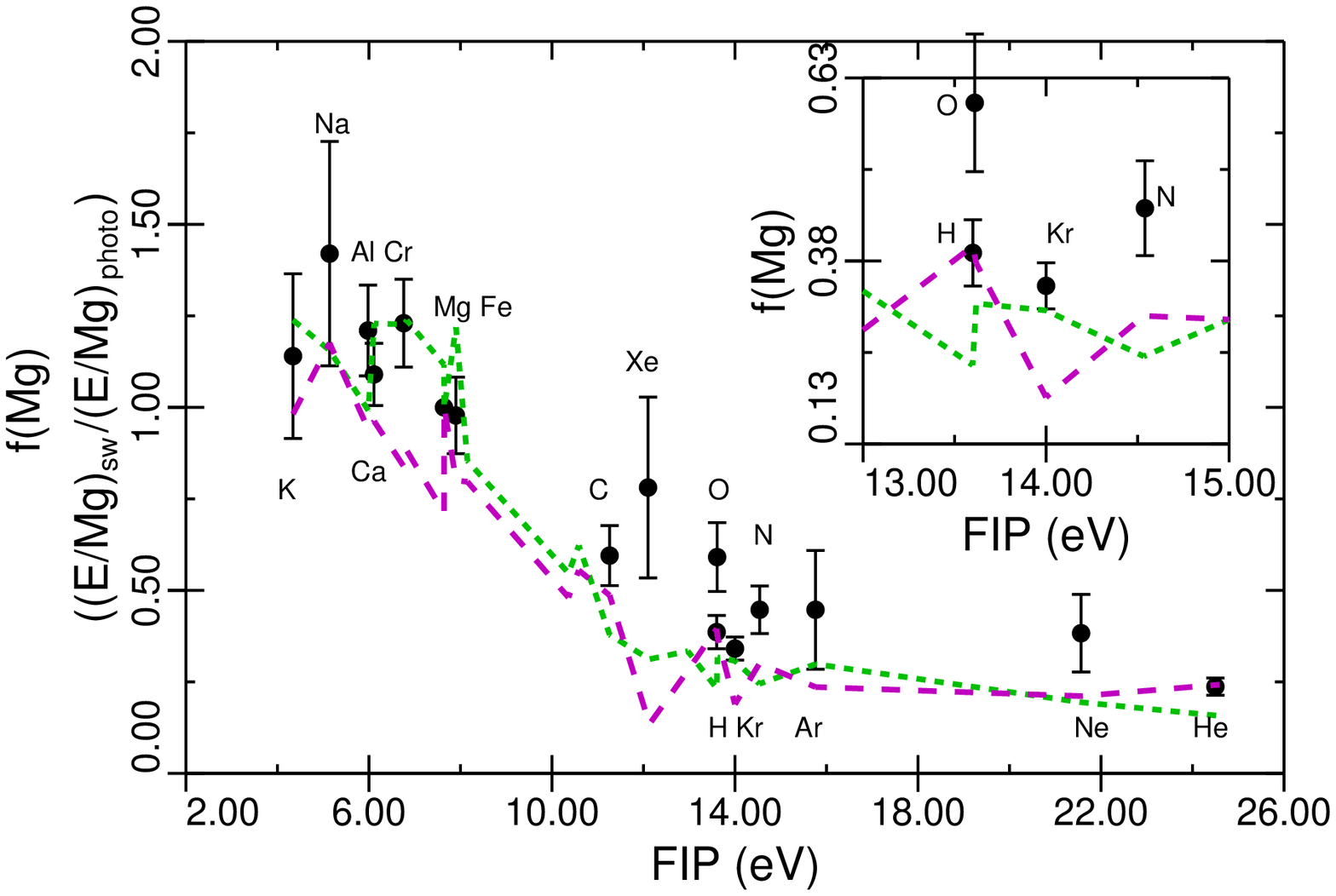}}
\caption{\label{fig4} Modeled fractionation patterns for Models 1 (left; low MDF)
and 2 (right; high MDF) from Table 1.
In each plot, the short dashed green line shows fractionation due to the
ponderomotive acceleration alone, and long dashed purple curve shows the effect of
ponderomotive acceleration and adiabatic invariant conservation. Symbols with error
bars show results from Genesis data analysis. Model 2 assumes a higher mass dependent
fractionation from the adiabatic invariant conservation. Model results for Kr and Xe assume
the same ionization balance as for Ar.}
\end{figure*}

\begin{table}
\begin{center}
\caption{Solar N and O Isotopic Abundances}
\begin{tabular}{|c|c|c|cc|}
\hline
Ratio & Model 1  & Model 2 & \multicolumn{2}{c}{Observations}\\
      & (low MDF)   & (high MDF) &  & \\
\hline  
$^{16}$O/$^{18}$O$^1$& 0.8 - 0.9 & 1.57 - 1.62& 2.2$^3$ &3.2$^4$\\
$^{25}$Mg/$^{26}$Mg$^1$& 0.5 - 0.8& 1.14 - 1.40& $\simeq 1$ $^5$& \\
$^{14}$N/$^{15}$N$^1$& 0.8 - 1.0& 1.63 - 1.68& $^6$& \\
$^{14}$N/$^{15}$N$^2$& 455$^7$& 452$^7$& 400 - 714 & $> 500^8$\\
\hline
\end{tabular}

\end{center}
{\tablecomments{$^1$Fractionation of bulk solar wind relative to photsphere,
\%/amu; light isotope enriched; $^2$ Absolute ratio; $^3$ data from McKeegan
et al. (2011) from Genesis; $^4$ data from Ayres et al. (2013) from
spectroscopy; $^5$ Heber et al. (2012b); $^6$ No directly measured
photospheric ratio; $^7$Calculated from our fractionations, used to correct
the Genesis measured solar wind $^{14}$N/$^{15}$N from Marty et al. (2011);
$^8$data from Fletcher et al. (2014) for Jupiter and Saturn respectively.}
\label{tab2}}
\end{table}

Our models are based on fractionations relative to \citet{asplund09} as
observed with ACE by \citet{pilleri15}. The most accurate Genesis data are
for Ca, Mg, Fe, H, and He, and we have emphasized the match to these in
tuning our models. There are no true photospheric abundances for Ar and Ne;
Kr is accurate, but is based on an interpolated CI chondrite solar abundance.
As noted, the adiabatic invariant model is only slightly below the low FIP
(+C) data.  The model agrees well with the high FIP H and He (plus Kr) data;
it is distinctly below the O and N points.

The upward displacement of the O and N fractionations above the model curves
in Fig. 3 may indicate that the photospheric abundances assumed for these
elements are too small. The latest revision of CNO photospheric abundances
\citep{asplund09,grevesse15,scott15a,scott15b} has recently been challenged
by \citet{vonsteiger16}, who argue that fast solar wind from polar coronal
holes is unfractionated and can be used to determine solar metallicity.  A
solar model based on this composition \citep{vagnozzi17} has been criticized
by \citet{serenelli16}. Although fast wind from polar coronal holes can be
considerably less fractionated than the fast wind seen in the ecliptic by
Genesis, a complete absence of FIP effect is not always supported by coronal
hole models of FIP fractionation, \citep{laming12,laming15}. However the
application of our FIP models to the Genesis data analyzed to date supports
the conclusion of \citet{vonsteiger16}, and is also more consistent with
higher values obtained previously by \citet{caffau08}, or even earlier by
\citet{grevesse98}.

The minimum amount by which the O abundance should increase to bring the
error bar into contact with the model is 0.06-0.14 dex (for Models 1 and 2
respectively), which moves the abundance from 8.69 of \citet{asplund09} to
8.75 - 8.83, in better agreement with \citet{caffau08} and/or
\citet{grevesse98}. For comparison, \citet{ayres13} give an O abundance of
8.75, and more recently \citet{cubasarmas17} give $8.86\pm 0.04$, both based
on spectroscopy.

The error bars on the Genesis data are one sigma, thus it is important to
await further analyses, especially of low and high speed regime samples. The
model result is driven by the fast wind model, for which the fractionation
ratio O/H $<1$ (see Figure 1b), but this is fundamentally a polar coronal
hole model applied to fast wind observed in the ecliptic. Measurements of the
slow wind abundance ratio O/H would remove this uncertainty. The possibility
exists once this is done of achieving a rather complete assessment of the
elemental and isotopic composition of the solar photosphere as a proxy for
the pre-solar nebula, {\em by methods completely independent} of those
employed to date.

\acknowledgements This work was supported by grants from the NASA
Heliophysics Supporting Research (NNH16AC39I) and Laboratory Analysis of
Returned Samples Programs (NNH17AE60I, NNH15AZ67I, NNX15AG19G) and by basic
research funds of the Chief of Naval Research. We are grateful to Yuan-Kuen
Ko for discussions of ACE data.

\end{document}